\documentclass[%
superscriptaddress,
twocolumn,
showpacs,
amsmath,amssymb,
aps,
prl,
floatfix,
showkeys,
reprint,
]{revtex4-1}

\usepackage{graphicx}
\usepackage{dcolumn}
\usepackage{bm}
\usepackage{amsmath}
\usepackage{amssymb}
\usepackage{multirow}
\usepackage{color}

\usepackage{booktabs}
\usepackage{tabularx}
\usepackage[english]{babel}
\usepackage[utf8]{inputenc}

\usepackage{enumerate}
\usepackage[cmyk,dvipsnames]{xcolor}

\definecolor{pacificb}{HTML}{1CA9C9}

\usepackage[normalem]{ulem}

\begin{document}

\title{Optimal protocol for spin-orbit torque switching of a perpendicular nanomagnet}

\author{Sergei M. Vlasov}
\thanks{These authors contributed equally to this work}
\affiliation{ITMO University, 197101 St. Petersburg, Russia}

\author{Grzegorz J. Kwiatkowski}
\thanks{These authors contributed equally to this work}
\affiliation{Science Institute of the University of Iceland, 107 Reykjav\'ik, Iceland}

\author{Igor S. Lobanov}
\affiliation{ITMO University, 197101 St. Petersburg, Russia}

\author{Valery M. Uzdin}
\affiliation{ITMO University, 197101 St. Petersburg, Russia}

\author{Pavel F. Bessarab}
\email[Corresponding author: ]{bessarab@hi.is}
\affiliation{ITMO University, 197101 St. Petersburg, Russia}
\affiliation{Science Institute of the University of Iceland, 107 Reykjav\'ik, Iceland}

\begin{abstract}
It is demonstrated by means of the optimal control theory that the energy cost of the spin-orbit torque induced reversal of a nanomagnet with perpendicular anisotropy can be strongly reduced by proper shaping of both in-plane components of the current pulse. The time-dependence of the optimal switching pulse that minimizes the energy cost associated with Joule heating is derived analytically in terms of the required reversal time and material properties. The optimal reversal time providing a tradeoff between the switching speed and energy efficiency is obtained. A sweet-spot balance between the field-like and damping-like components of the spin-orbit torque is discovered; it permits for a particularly efficient switching by a down-chirped rotating current pulse whose duration does not need to be adjusted precisely. 
\end{abstract}

\maketitle

The discovery of spin-orbit torque (SOT) is a notable milestone in the development of spintronics as it 
has made it possible to boost the efficiency of electrical manipulation of magnetism 
compared to conventional spin transfer torques~\cite{gambardella2011current, miron2011perpendicular, liu2012spin}. Magnetization switching by current-induced field-like (FL) and damping-like (DL) components of SOT is a particularly important application providing a basis for low-power bit operations in nonvolatile technologies~\cite{aradhya2016nanosecond, garello2014ultrafast}. 

Typically, a SOT-induced magnetization reversal is realized by applying an in-plane current in a heavy-metal (HM) layer on which a switchable ferromagnetic (FM) element is placed. Elements with perpendicular magnetic anisotropy (PMA) are under a special focus due to their technological relevance. In a conventional protocol involving a one-dimensional direct current, the deterministic reversal of the PMA element relies only on the DL SOT~\cite{lee2013threshold,fukami_2016}. As a result, the switching current density and thereby the energy cost of switching are not as low as they could possibly be if the FL SOT was also used~\cite{taniguchi_2015}. FL SOT-induced switching can also be realized, but this requires a precise control over the current pulse duration so as to avoid back-switching~\cite{lee2018oscillatory}. Moreover, to achieve definite switching of the PMA element, some symmetry breaking needs to be established 
either by applying an external magnetic field~\cite{miron2011perpendicular, liu2012spin, avci2014fieldlike}, which can be mimicked by exchange coupling with an additional magnetic layer~\cite{lau2016spin, van2016field, fukami2016magnetization,oh_2016}, or by introducing lateral asymmetry~\cite{yu2014switching, yu2014magnetization} or tilted anisotropy~\cite{torrejon2015current, you2015switching}. Out-of-plane torque for field-free switching can also be generated due to the low-symmetry point groups at the HM/FM interfaces~\cite{liu_2021} or thanks to the out-of-plane polarization of spin currents generated in an additional magnetic layer~\cite{baek_2018}.

Note that the complications associated with the conventional SOT-induced magnetization reversal originate from use of one-dimensional direct current. Such a simple switching pulse provides a limited control over switching and makes it hard to realize full potential of SOT. 

The issues arising in SOT-induced 
switching can be solved by proper shaping of the current pulse. Pulse optimization has been studied extensively in the past in the context of magnetization reversal by means of applied magnetic field~\cite{thirion2003switching,sun_2006,rivkin_2006,sun2006theoretical,
woltersdorf_2007,
bertotti_2009,
barros2011optimal,barros2013microwave,cai2013reversal} and spin-transfer torque~\cite{wang_2007,tretiakov_2010}. 
For SOT-induced reversals, 
this approach remains 
unexplored,  
although its potential has recently been demonstrated 
by Zhang {\it et al.}~\cite{zhang_2018} who proposed to use both in-plane components of the current to realize field-free switching of a PMA element; Assuming a fixed magnitude but variable direction of the current, they 
obtained a strong reduction in the switching current density and derived a pulse yielding the shortest switching time. However, constraints imposed on the current pulse prevented previous studies from identifying the theoretical minimum of the energy cost of SOT-induced switching.
Optimization of the switching protocols with respect to materials properties and identification of the right balance between switching speed and energy-efficiency have also been missing so far despite the great fundamental and technological importance of this analysis. Moreover, it is still unclear whether the protocols involving 2D currents are stable enough against thermal fluctuations to be realized in practice. 

In this Letter, 
we identify energy-efficient switching pulses 
by applying a systematic approach based on the optimal control theory.  
To make most of the pulse optimization, we do not apply any constraints on the pulse shape and consider independent variations of both in-plane components of the current. 
We obtain a complete analytical solution for optimal control paths (OCPs) of field-free magnetization switching, i.e. the reversal trajectories minimizing the energy cost associated with Joule heating, and 
derive optimal switching pulses from the obtained OCPs. 

Our analytical solution provides a theoretical limit for energy-efficient control of SOT-induced magnetization switching and reveals noteworthy exact results connecting the minimum energy cost, optimal switching current and switching time with relevant materials properties. We uncover 
a previously overlooked sweet-spot ratio of the FL and DL torques for which a particularly appealing switching protocol is possible. It is robust, corresponds to the lowest energy cost, shortest switching time and 
involves quite simple switching pulse that can likely be realized in the laboratory.

\begin{figure}[!ht]
\centering
\includegraphics[width=\columnwidth]{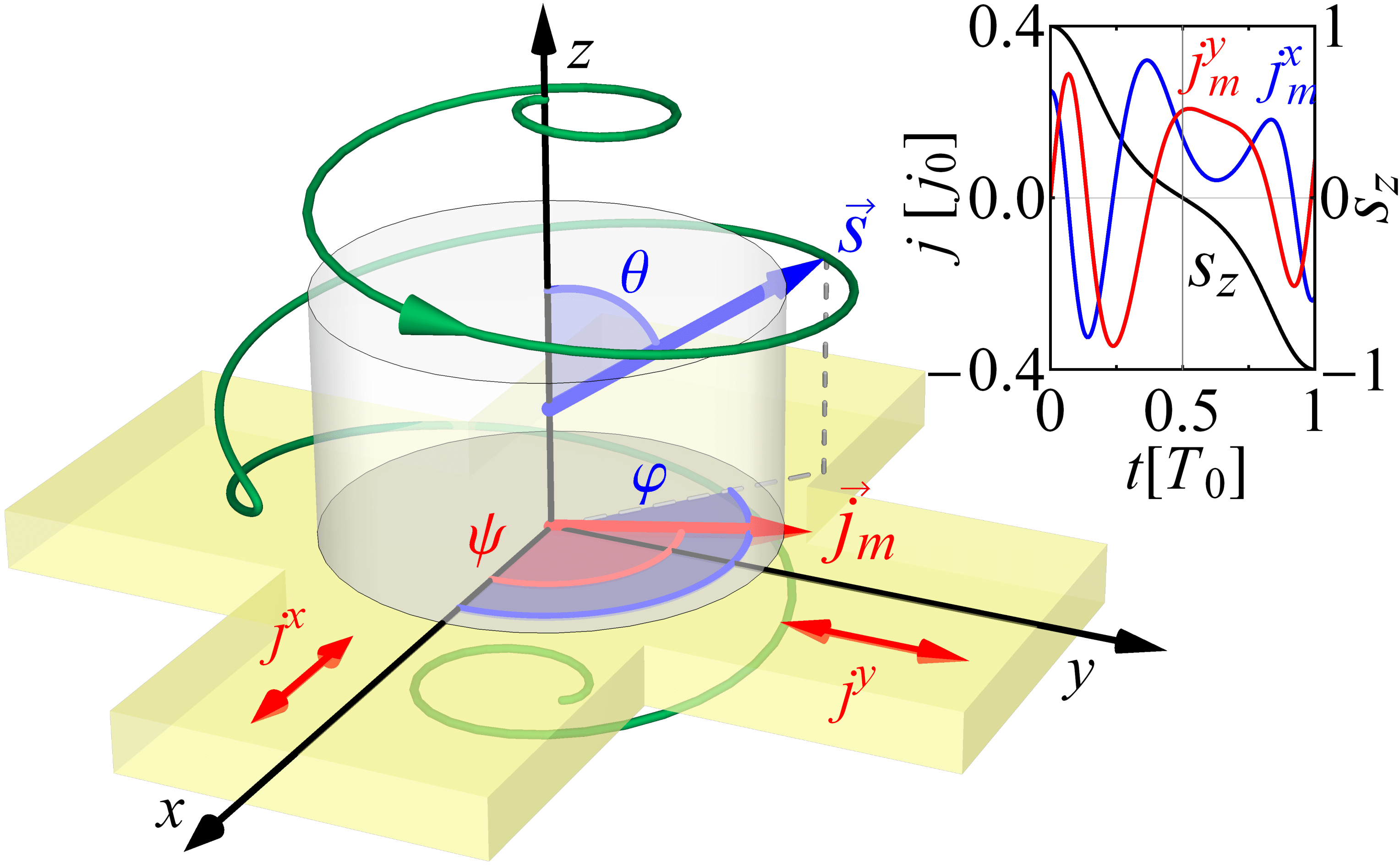}
\caption{\label{fig:1} Energy-efficient switching of a PMA-nanoelement (cylinder) by an optimal 2D electric current pulse $\vec{j}_m$ flowing in the heavy-metal substrate (cross). 
The calculated optimal control path for the switching for $\alpha=0.1$ and $\xi_D=3.56 \xi_F$ 
is shown with the green line. 
The direction of the normalized magnetic moment $\vec{s}$ of the element (optimal current $\vec{j}_m$) is shown with the blue (red) arrow. The inset shows the time-dependence of $x$- and $y$-components of $\vec{j}_m$, and $z$-component of $\vec{s}$.
}
\end{figure}

Figure~\ref{fig:1} shows the simulated PMA element whose magnetic moment is reversed by an in-plane current via SOT. The element is assumed small enough to be treated essentially as a single-domain particle at any time of the reversal process. 
The energy $E$ of the system is defined by the anisotropy along $z$ axis, 
\begin{equation}
    \label{eq:energy}
    E = -Ks_z^2,
\end{equation}
where $s_z$ is a $z$-component of the normalized magnetic moment $\vec{s}$, and $K>0$ is the anisotropy constant. 
The task is to identify the optimal current pulse that reverses the magnetic moment from $s_z=1$ at $t=0$ to $s_z=-1$ at $t=T$, with $T$ being the switching time. 
Both the amplitude and the direction of the current $\vec{j}$ in the heavy-metal layer are allowed to vary in time. This 
can be realized in 
the cross-type geometry permitting independent control of 
both in-plane components of the current, 
see Fig.~\ref{fig:1}. 
The efficiency of the reversal is naturally defined by the amount of Joule heating generated in the resistive circuit during the switching process ~\cite{tretiakov_2010}. In particular, the optimal reversal is achieved when the cost functional 
\begin{equation}
\label{eq:cost}
    \Phi = \int_0^T |\vec{j}|^2dt,
\end{equation}
is minimized. This optimal control problem is subject to a constraint imposed by the zero-temperature Landau-Lifshitz-Gilbert equation describing the dynamics of the magnetic moment under SOT~\cite{haney2013current}: 
\begin{eqnarray}
\label{eq:eom}
    \dot{\vec{s}} = &-&\gamma\vec{s}\times\vec{b}
	+ \alpha\vec{s}\times\dot{\vec{s}} \nonumber\\
	&+& \gamma\xi_{F}\vec{s}\times(\vec{j}\times\vec{e}_z) 
	+ \gamma\xi_{D}\vec{s}\times\bigl[\vec{s}\times (\vec{j}\times\vec{e}_z)\bigr].
\end{eqnarray}
Here, $\gamma$ is the gyromagnetic ratio, $\alpha$ is the damping parameter, $\vec{e}_z$ is the unit vector along the $z$ axis, and $\vec{b}$ is the anisotropy field: $\vec{b} \equiv -\mu^{-1} \partial E / \partial \vec{s}$, with $\mu$ being the magnitude of the magnetic moment. The third and the fourth terms in the rhs of Eq.~(\ref{eq:eom}) represent FL and DL components of the SOT, respectively. The coefficients $\xi_F$ and $\xi_D$ are defined by the spin Hall angle, saturation magnetization and thickness of the FM element, as well as by dimensionless factors -- efficiencies -- characterizing the weights of the SOT components~\cite{yoon2017anomalous}. 

To find the optimal switching current $\vec{j}_m(t)$ that makes $\Phi$ minimum, we follow the paradigm we applied earlier to the magnetization reversal induced by applied magnetic field~\cite{kwiatkowski2021optimal}: The energy cost is first expressed in terms of the switching trajectory and then minimized so as to find the OCPs for the switching process; After that, the optimal switching pulse $\vec{j}_m(t)$ is derived from the OCPs. Qualitative difference between the field torque and SOT makes these calculations significantly more involved compared to Ref.~\cite{kwiatkowski2021optimal}.

We start by expressing the amplitude 
of $\vec{j}$ in terms of the dynamical trajectory using Eq.~(\ref{eq:eom}): 
\begin{equation}
    \label{eq:j}
    j =  \frac{2K}{\mu}\frac{(1+\alpha^2)\tau_0\dot{\theta} + \alpha \sin\theta\cos\theta}{(\alpha \xi_\text{D}-\xi_\text{F})\cos\varphi^\prime- (\alpha \xi_\text{F} + \xi_\text{D})\cos\theta\sin\varphi^\prime}.
\end{equation}
Here, $\tau_0=\mu(2K\gamma)^{-1}$ defines the timescale of Larmor precession, $\theta$ and $\varphi$ are the polar and azimuthal angles of $\vec{s}$, respectively, and $\varphi^\prime \equiv \varphi-\psi$, with $\psi$ being the angular coordinate of the current (see Fig.~\ref{fig:1}). It is clear from Eq.~(\ref{eq:j}) that for a given magnitude of the current-generated torque, overall increase in the SOT coupling coefficients leads to a proportional decrease in the switching current, and thereby the energy cost. To elucidate a nontrivial effect of the SOT parameters, we 
introduce a variable $\beta$ characterizing the balance between the SOT components, via the following parametrization: 
\begin{equation}
\xi_\text{F}=\xi\cos\beta,\quad \xi_\text{D}=\xi\sin\beta,\label{eq:param}
\end{equation}
where $\xi$ is the magnitude of the total SOT coupling, i.e. $\xi\equiv\sqrt{\xi_\text{D}^2+\xi_\text{F}^2}$.

On substituting Eq.~(\ref{eq:j}) into Eq.~(\ref{eq:cost}), the energy cost of the reversal becomes a functional of the switching trajectory. Taking into account Eq.~(\ref{eq:param}) and minimizing $\Phi$ with respect to $\varphi^\prime$, we obtain an optimal value of $\varphi$ with respect to $\psi$ (see also Ref.~\cite{zhang_2018}): 
\begin{equation}
 \varphi^\prime=\varphi-\psi = \arctan\left[\tan(\beta+\eta)\cos\theta\right],\label{eq:optimal_dir}
\end{equation}
where $\eta \equiv \arctan(\alpha)$. 
After eliminating the $\varphi^\prime$-dependence of $\Phi$ with the use of Eq.~(\ref{eq:optimal_dir}), the Euler-Lagrange equation for $\theta$ can be derived. It's solution satisfying the boundary conditions $\theta(0)=0$, $\theta(T)=\pi$ is expressed in terms of Jacobi elliptic functions~\cite{supplement,abramowitz1948handbook}. Finally, optimal $\varphi(t)$ is obtained from optimal $\theta(t)$ using the equation of motion [see Eq.~(\ref{eq:eom})], where $\psi$ is eliminated using Eq.~(\ref{eq:optimal_dir}). In this way, the OCP describing the switching trajectory that minimizes the energy cost is completely defined. 
It corresponds 
to the magnetic moment rotating steadily from the initial state minimum to the final one and at the same time precessing around the anisotropy axis. Depending on whether the sign of $\tan(\beta+\eta)$ is negative or positive, the sense of precession changes before or after the system crosses the energy barrier, respectively. However, the barrier crossing happens exactly at $t=T/2$, which results from a 
general symmetry of the solution $\theta_m(t)$: $\theta_m(T/2+t^\prime)=\pi-\theta_m(T/2-t^\prime)$, $0\le t^\prime \le T/2$.  
The OCP for $\alpha=0.1$, $\xi_D=3.56\xi_F$ and $T=T_0$ [see Eq. (\ref{eq:opt_sw_t})] is shown in Fig.~\ref{fig:1}.

Optimal control $\vec{j}_m(t)$ can be derived from the OCP using Eqs.~(\ref{eq:j}) and (\ref{eq:optimal_dir}). It corresponds to a rotating current whose sense of rotation changes together with that of the magnetic moment precession around the anisotropy axis. Although the form of $\vec{j}_m(t)$ is rather complex in a general case (see the inset in Fig.~\ref{fig:1}), the following properties hold exactly. For finite damping, the amplitude of the optimal current is modulated such that it reaches a maximum value exactly at $t=T/4$ and a minimum value at $t=3T/4$, with the difference between the extremal values given by %
\begin{equation}
\label{eq:deltaj}
    \Delta j_\text{m} = \frac{4 \alpha j_0}{\sqrt{1+\alpha^2}\left(|\cos(\beta+\eta)|+1\right)},
\end{equation}
where $j_0=K(\mu\xi)^{-1}$. Equation (\ref{eq:deltaj}) particularly signifies that the current amplitude is constant for zero damping. 
The average current amplitude can also be expressed analytically:
\begin{equation}
\label{eq:average}
    \langle j_\text{m}\rangle = \frac{4j_0\tau_0\sqrt{1+\alpha^2}\mathcal{K}\left[\sin^2(\beta+\eta)\right]}{T},
\end{equation}
where $\mathcal{K}[.]$ is the complete elliptic integral of the first kind~\cite{supplement,abramowitz1948handbook}. Notably, the average current does not depend on the height of the energy barrier defined by the magnetic anisotropy. 
Additionally, the following symmetry holds in general: $j_\text{m}(0)=j_\text{m}(T/2)=j_\text{m}(T)$. 

The minimum energy cost $\Phi_\text{m}$ is a monotonically decreasing function of the switching time exhibiting two asymptotic regimes. For fast switching, $\Phi_\text{m}(T)$ scales inversely with $T$ and does not depend of the magnetic potential: 
\begin{equation}
\label{eq:fast}
    \Phi_\text{m} \approx \frac{4(1+\alpha^2)\mathcal{K}^2\left(\sin^2(\beta+\eta)\right)}{T\gamma^2\xi^2}, \quad T\ll (\alpha+1/\alpha)\tau_0.
\end{equation}
At infinitely long switching time, $\Phi_m$ approaches the lower limit: 
\begin{equation}
\label{eq:slow}
    \Phi_\text{m} \rightarrow \frac{4\alpha K\log\left[1+\tan^2(\beta+\eta)\right]}{\mu\gamma\xi^2\sin^2(\beta+\eta)}, \quad T\rightarrow \infty.
\end{equation}
Intersection of the asymptotics gives a characteristic switching time 
providing a tradeoff between the switching speed and energy efficiency.  
For a given T, the energy cost corresponding to the constant-amplitude pulse derived in~\cite{zhang_2018} always exceeds $\Phi_m(T)$ and even diverges at $T\rightarrow\infty$~\cite{supplement}.

\begin{figure}[!ht]
\centering
\includegraphics[width=1.0\columnwidth]{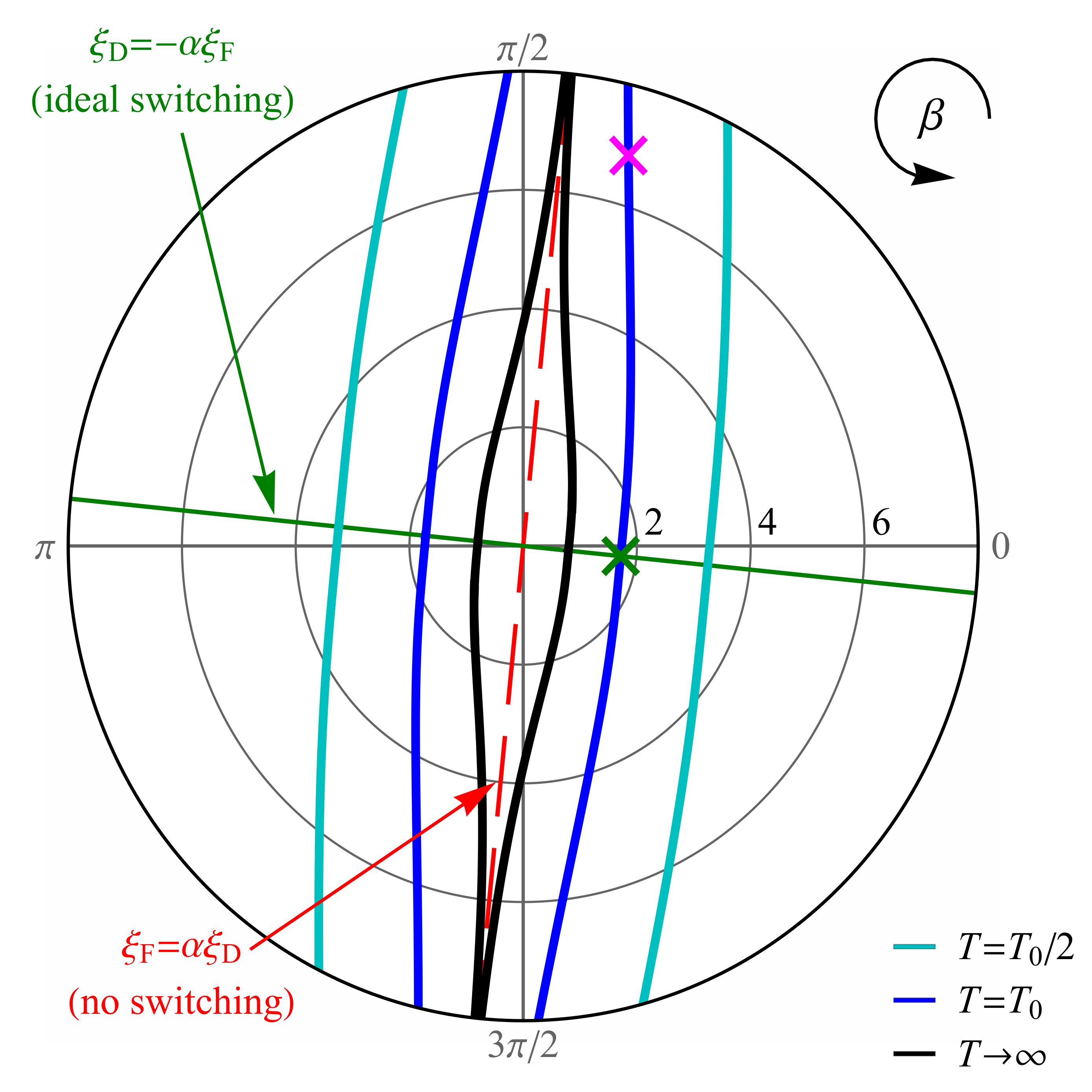}
\caption{\label{fig:2} Minimum energy cost of magnetization switching in units of $j_0^2 \tau_0$ as a function of $\beta$ for $\alpha=0.1$ and several values of the switching time. 
Green solid line indicates the ideal ratio of the SOT coefficients, red dashed line marks the ratio that prohibits switching. The green (magenta) cross indicates ideal (non-ideal) parameter values for which the optimal current pulse is shown in Fig.~\ref{fig:3}(a) (inset of Fig.~\ref{fig:1}).
}
\end{figure} 

$\Phi_\text{m}$ as a function of $\beta$ is shown in Fig.~\ref{fig:2}.  
$\Phi_\text{m}$ diverges when $\alpha\xi_\text{D}=\xi_\text{F}$ signifying 
no switching 
for this 
ratio of the SOT coefficients, which was also pointed out in Ref.~\cite{zhang_2018}. The divergence originates from the vanishing torque in the direction of increasing $\theta$ at the equator. To realize 
switching in this case, an additional force such as external magnetic field needs to be applied to the system, similar to what is done in conventional SOT-induced switching~\cite{fukami2016magnetization}. 

On the other hand, $\Phi_\text{m}(\beta)$ has minima at $\beta=\beta^\ast$, with 
$\beta^\ast$ 
defined by $\tan(\beta^\ast+\eta)=0$. 
The minima correspond to an ideal ratio between the SOT coefficients, 
\begin{equation}
    \label{eq:optimal}
    \xi_\text{D}=-\alpha\xi_\text{F},
\end{equation}
for which the switching is particularly efficient. 
For this 
ratio, the torque generated 
by $\vec{j}_m$ is invested entirely into the 
increase in $\theta$, i.e. only into the motion which is relevant for switching. 
Such a rational use of an external stimulus can always be achieved for the optimal magnetization reversal induced by applied magnetic field~\cite{kwiatkowski2021optimal} via the adjustment of all three components of the 
field. For SOT-induced switching, realization of the SOT exclusively in the direction of increasing $\theta$ 
is only possible for the ideal balance between $\xi_\text{D}$ and $\xi_\text{F}$ due to the confinement of the switching current to $xy$-plane.

The characteristic switching time defined by crossing of the asymptotics of $\Phi_m$ [see Eqs.~(\ref{eq:fast}) and (\ref{eq:slow})] also approaches the minimum value (with respect to the variation of $\beta$) $T_0$ for the ideal ratio of the SOT coefficients:
\begin{equation}
    \label{eq:opt_sw_t}
    T_0 = \frac{(1+\alpha^2)\pi^2}{2\alpha}\tau_0. 
\end{equation}
For $\alpha=0.1$, $T_0$ corresponds to just a few oscillations of the magnetic moment. Upon substituting $\beta=\beta^\ast$ and $T=T_0$ into Eq.~(\ref{eq:average}), the average switching current becomes (here and below, an asterisk signifies a quantity corresponding to the ideal ratio between the SOT coefficients):
\begin{equation}
    \langle j_m^\ast\rangle = \frac{4\alpha j_0}{\pi\sqrt{1+\alpha^2}}.
\end{equation}
Noteworthy, this characteristic current scales with $\alpha$ 
in the low damping regime making 
it significantly smaller than the critical current for conventional SOT-induced switching of a PMA-element~\cite{lee2013threshold}.  

\begin{figure}[!ht]
\centering
\includegraphics[width=1.0\columnwidth]{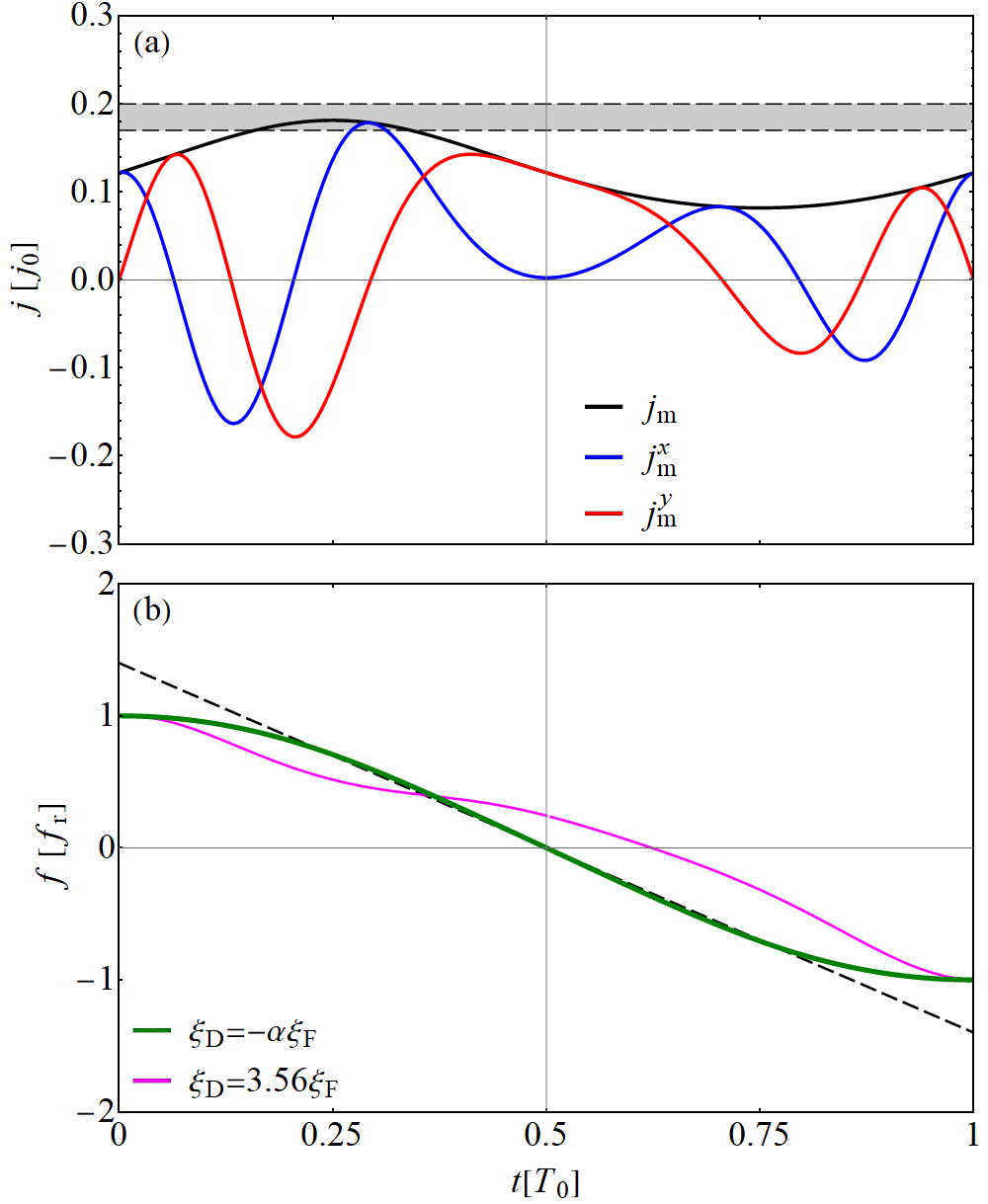}
\caption{\label{fig:3} (a) Time-dependence of the optimal switching current for the ideal ratio of the SOT coefficients [see Eq.~(\ref{eq:optimal})], $\alpha=0.1$ and $T=T_0$. Gray area in between the dashed lines indicates the range of amplitudes used for the simplified current pulse [see Eq.~(\ref{eq:js})]. 
(b) Time-dependence of the frequency of the optimal current pulse for the ideal (solid green line) and non-ideal (solid magenta line) ratio of the SOT coefficients. Dashed black line shows the frequency variation for the simplified pulse.
}
\end{figure}

Together with the shortest switching time and improved energy efficiency, the ideal ratio of the SOT coefficients provides particularly simple switching protocol. Indeed, the Euler-Lagrange equation describing the OCP simplifies significantly and becomes identical to that derived for the magnetic field-induced switching~\cite{kwiatkowski2021optimal}. 
The optimal switching pulse gets simpler as well. Its rotation frequency becomes equal to the resonant frequency of the system:
\begin{equation}
    f_\text{m}^\ast \equiv \frac{1}{2\pi}\dot{\psi}_m^\ast= \frac{\cos(\theta_m^\ast)}{2\pi\tau_0(1+\alpha^2)}.
\end{equation}
The frequency decreases monotonically with time and exhibits a symmetry: $f_\text{m}^\ast(T/2+t^\prime)=-f_\text{m}^\ast(T/2-t^\prime)$, $0\le t^\prime \le T/2$, signifying the sign change exactly at $t=T/2$. Time dependence for the switching current and its frequency for~$\alpha=0.1$, $T=T_0$ and $\xi_D=-\alpha\xi_F$ is shown in Fig.~\ref{fig:3}. Overall, the switching current for the ideal balance between the SOT coefficients resembles a down-chirped pulse whose rotation reverses at the instant of the barrier crossing, and amplitude is fairly constant in the low damping regime [see Eq.~(\ref{eq:deltaj})]. 

Motivated by this result, we further investigate whether the substitution of the optimal control by a simplified pulse represented by a rotating current with constant amplitude and time-linear frequency sweep indeed leads to magnetization switching in the PMA element. For this, we simulate the magnetization dynamics induced by current $\vec{j}_s(t)$ given by the following ansatz: 
\begin{equation}
    \vec{j}_s(t) = j_s\cos\bigl[\Omega(t)\bigr]\vec{e}_x+j_s\sin\bigl[\Omega(t)\bigr]\vec{e}_y,\label{eq:js}
\end{equation}
where $\Omega(t)=2\pi f_\text{max}\left(t-t^2/T\right)$. 
We additionally include thermal noise in the simulations to verify robustness of the switching~\cite{supplement}. We perform the simulations for $T=T_0$, $\alpha=0.1$, and the ideal ratio of the SOT coefficients. We find that $\vec{j}_s(t)$ reliably induces magnetization switching as long as its amplitude $j_s$ is large enough and its initial frequency $f_\text{max}$ slightly exceeds the resonant frequency $f_r$ at the energy minimum, $f_r\equiv[2\pi\tau_0(1+\alpha^2)]^{-1}$. In particular, for $f_\text{max}=1.4f_r$ and a typical for a memory element thermal stability factor~\footnote{Thermal stability factor is defined as a ratio between the energy barrier and the thermal energy.} of 60 the switching probability increases from 0.89 to 0.97 as $j_s$ changes from $0.17 j_0$ to $0.18 j_0$, and becomes practically unity for $j_s=0.2j_0$. 
As soon as the moment roughly reverses its orientation at $t\approx T$, the pulse can be terminated, but extending the pulse duration beyond $T$ does not compromise switching, as confirmed by our simulations. The absense of unwanted instabilities such as back switching is expected since interaction of $\vec{j}_s(t)$ with the magnetic moment becomes progressively less effective for $t>T$ where the pulse frequency exceeds the resonant frequency of the system, and the moment stays locked in the reversed position. 
The switching protocol produced by the rotating current does not require fine tuning of the pulse duration and is therefore robust. 

On the other hand, the simplified switching protocol does require the ratio between the SOT coefficients to be close enough to the ideal value. 
Otherwise, the optimal switching pulse can be quite different from that described by Eq.~(\ref{eq:js}) [see 
the inset of Fig.~\ref{fig:1} and 
Fig.~\ref{fig:3}(b)], 
and no stable switching on the timescale of moment oscillations can be obtained using the simplified pulse~\cite{supplement}, although slower reversal involving multiple precession motion can still be achieved regardless the ratio of the SOT coefficients~\footnote{As long as $\alpha\xi_\text{D}\neq\xi_\text{F}$.} thanks to the autoresonant excitation~\cite{go_2017,klughertz_2014,klughertz_2015}. It is however expected that slow autoresonant switching is quite sensitive to thermal fluctuations that tend to disturb the phase locking.  

Experimental realization of fast and energy-efficient switching of PMA elements by means of chirped rotating current requires the SOT coefficients to have opposite signs, with FL torque being significantly larger than DL torque [see Eq. (\ref{eq:optimal})]. Several systems where this scenario is realized have been reported, see e.g.~\cite{kim2013,qiu2014,ramaswamy2016} and Table II in~\cite{manchon2019}. Moreover, the torques can be tuned by inducing piezoelectric strain~\cite{filianina_2020} or generating orbital currents~\cite{ding_2020}.

While only the FL and DL torques are considered here, the SOT can have additional angular dependence~\cite{garello_2013}. Optimization of the magnetization reversal for such an extended SOT model is a challenging problem that goes beyond the scope of the present study. From general arguments it follows that further decrease in the energy cost is conceivable in the extended model, but this remains to be explored. Nevertheless, some arguments on how the combination of the SOT coefficients in the extended model affects the magnetization switching can be provided without actually solving the optimal control problem~\cite{supplement}.

In conclusion, we have presented a theoretical limit for the minimal energy cost of the SOT-induced magnetization reversal in the PMA nanoelement 
at zero applied magnetic field and derived corresponding optimal switching current pulse as a function of the reversal time and relevant material properties. We have identified an ideal ratio of the SOT coefficients corresponding to a particularly efficient, robust and simple switching protocol. The average switching current for the ideal balance between the DL and FL torques scales with the Gilbert damping parameter which makes it significantly lower than a critical current in conventional switching protocols. Our results reveal a target for the design of PMA systems for energy-efficient applications and 
inspire experimental studies of pulse shaping for the optimization of the current-induced magnetization dynamics and switching.

\begin{acknowledgments}
The authors would like to thank T. Sigurj\'onsd\'ottir and C. Back for helpful discussions. This work was funded by the Russian Science Foundation (Grant No. 19-72-10138), the Icelandic Research Fund (Grant No. 184949), and the University of Iceland Research Fund (Grant No. 15673).  
\end{acknowledgments}

%
\end{document}